# Imaging with an Almost Perfect Lens

## R. Merlin

*FOCUS Center and Department of Physics*
*The University of Michigan, Ann Arbor, MI 48109-1120*

PACS numbers: 78.20.Ci, 42.30.-d, 41.20.Jb, 78.66.-w

## ABSTRACT

The problem of imaging for a nearly-perfect lens, namely, a slab of a left-handed material with refractive index $n = -(1-\sigma)^{1/2}$ is solved analytically for $|\sigma| \ll 1$. The electromagnetic field behavior is determined largely by singularities arising from the excitation of surface polaritons with wavevector $q \rightarrow \infty$. Depending on the sign of $\sigma$, the near field is either odd or even with respect to the lens middle plane. Images exhibit an anomalous interference pattern with length scale determined by the width of the slab. Consistent with recent studies by Smith et al. [Appl. Phys. Lett. **82**, 1506 (2003)] and Gómez-Santos [Phys. Rev. Lett. **90**, 077401 (2003)], the resolution depends logarithmically on $|\sigma|$.

Abbe proved in the 1870s [1] that the smallest feature a lens can image is limited by diffraction to $\sim \lambda/2n$ where $\lambda$ is the wavelength of light and $n$ is the refractive index. Despite many attempts to circumvent this barrier involving, in particular, nonlinear [2], solid-immersion [3], and other mainly near-field [4] techniques, significant progress has remained elusive. Recently, Pendry [5] argued that a slab of a left-handed (LH) substance with $\varepsilon=\mu=-1$ should behave as a perfect lens ($\varepsilon$ and $\mu$ are, respectively, the permittivity and the magnetic permeability). The terms optical left- and right-handedness were introduced by Veselago [6] in the 1960s to distinguish substances with both $\varepsilon < 0$ and $\mu < 0$ and, thus, $n < 0$ from conventional, right-handed (RH) $n > 0$ media for which $\varepsilon$ and $\mu$ cannot be both negative at a given frequency. Following Pendry's work [5] and the experimental demonstration of negative refraction at microwave frequencies [7], LH substances have attracted a great deal of interest along with some contention [8-27]. While recent experiments [26-27] seem to have put to rest concerns regarding the far field behavior of negative-refraction slabs, the question of near-field focusing has remained highly controversial [17-25] primarily because of the divergences plaguing the theory at $\varepsilon=\mu=-1$. Here, we provide an analytical answer to this problem. Our results support Pendry's concept of a perfect lens while raising new questions about the feasibility of the proposal.

We consider the propagation of electromagnetic waves from vacuum to an LH-medium occupying the half space $z > 0$, and we assume that $\text{Im}(\varepsilon)= \text{Im}(\mu) = 0$. The case $\varepsilon=\mu=-1$ [5] will be referred to as ideal refraction. Let **H** and **E** be the magnetic and the electric field, and $\omega$ the frequency of light. The transverse magnetic solutions to Maxwell's equations, the so-called $p$–polarized waves, are of the form $H_y = h(z)\exp(-i\omega t + iqx)$, $H_x = H_z = 0$ (with few modifications, primarily replacing $\varepsilon$ by $\mu$ and **H** by **E**, arguments similar to those discussed below apply as well to transverse-electric or $s$–modes). From the expression for **H**, we can obtain the electric field using $\mathbf{E} = -(ic/\varepsilon\omega)\nabla\times\mathbf{H}$. For $z > 0$, we have $h = M^+\exp(+\kappa z) + M^-\exp(-\kappa z)$ where

$$\kappa = \begin{cases} i\sqrt{\varepsilon\mu\omega^2/c^2 - q^2} & q^2 < \varepsilon\mu\omega^2 \\ \sqrt{q^2 - \varepsilon\mu\omega^2/c^2} & q^2 > \varepsilon\mu\omega^2 \end{cases} \qquad (1)$$

while, for vacuum, $h = A^+\exp(+\kappa_0 z) + A^-\exp(-\kappa_0 z)$ with $\kappa_0 = \kappa(\varepsilon=\mu=1)$. We observe that $\kappa = \kappa_0$ for $\varepsilon = \mu = -1$ and also that, since $H_y$ and $(\partial H_y/\partial z)/\varepsilon$ must be continuous at the boundary, $A^- = M^+$ and $A^+ = M^-$ for an ideal interface. Hence, refraction causes a reversal in the sign of the exponent for both propagating ($q^2 < \varepsilon\mu\omega^2$) and evanescent ($q^2 > \varepsilon\mu\omega^2$) waves. In a slightly modified form, this feature accounts for the unusual optical properties of LH substances and, in particular, for the remarkable converging lens performance of planar RH/LH interfaces [6]. The latter effect can be understood by considering a two-dimensional source at $z = -\ell$ for which the radiative component in vacuum can be generally written as

$$H_y^R = \int_{-\omega/c}^{+\omega/c} H(q) e^{iqx + i\sqrt{\omega^2/c^2 - q^2}|z+\ell|} dq \; . \qquad (2)$$

Then, for an ideal interface (2) is also the solution for $z < 0$. For $z > 0$, we readily obtain

$$H_y^R = \int_{-\omega/c}^{+\omega/c} H(q) e^{iqx - i\sqrt{\omega^2/c^2 - q^2}(z-\ell)} dq \qquad (3)$$

which exhibits aberration-free focusing at $z = \ell$. The focal point arises because, after crossing $z = 0$, the waves begin to lose at exactly the same rate the phase they had gained in their motion in vacuum. As first discussed by Veselago [6], the ideal vacuum-LH interface is but a particular case of the problem of refraction at an RH/LH boundary. Veselago [6] showed that LH materials generally behave as optical media with negative refractive index $n_L = -(\varepsilon\mu)^{1/2}$ so that a flat interface connecting such a medium to an RH substance, with refractive index $n_R$, acts as a converging lens with focal length given by $n_L \ell/(n_L - n_R)$ (images are free of aberrations only for the ideal case $n_L/n_R = -1$).

The above results apply only to radiative modes and, thus, to length scales $\geq \lambda$. Features of smaller sizes are contained in the near-field [5]

$$H_y^{NF} = \int_{|q|>\omega/c} H(q) e^{iqx - \sqrt{\omega^2/c^2 - q^2}|z+\ell|} dq \quad . \tag{4}$$

Because evanescent waves cannot be amplified in conventional refraction (this can be attained in some sense with mirrors), the dimensions of the focal spot are at best of order $\lambda$ [1]. Yet, for ideal RH-LH refraction amplification seems possible given that $\exp(-\kappa_0 z)$ connects to $\exp(\kappa_0 z)$ for $A^- = M^+$. Thus, one might be led to believe that evanescent modes too focus at $z = \ell$ and, therefore, that a perfectly resolved image can be obtained.

However, it is immediately obvious that this argument poses a problem since physically sound solutions cannot grow away from the interface. As pointed out by Haldane [22] and others [23-25], the absence of a well-behaved solution is due to resonant excitation of surface plasmons or, more generally, polaritons causing the field to become infinitely large at $\varepsilon = -1$ (this problem does not affect the far field). The dispersion of these modes obeys $\kappa/\kappa_0 = -\varepsilon$ [28,29] and, thus, the frequency at which $\varepsilon = -1$ is always the solution for $q \to \infty$ where the density of states diverges. We observe that this singularity can be avoided by adding a dissipative term, and that other approaches for introducing a *q*-cutoff have been proposed [22,23].

The considerations for a single boundary can be easily extended to two interfaces and, in particular, for a negative-refraction slab occupying the region $0 < z < d$ and sandwiched by vacuum. With the source, as before, at $z = -\ell$ and provided $d > \ell$, it can be shown that there are now two far-field images which are aberration-free for $\varepsilon = \mu = -1$. The first image is inside the medium, at $z = \ell$, and the second one is at $z = 2d-\ell$ [6]. Notably, and different from the single interface, the slab geometry admits an acceptable solution for evanescent modes at $\varepsilon = \mu = -1$ since the exponential that grows with *z* inside the slab can be matched to a decaying exponential outside. This is the celebrated Pendry's solution which leads to a perfect image of the source, with infinite resolution, at $z = 2d-\ell$ [5]. The evanescent modes also converge to this point because, after decaying from $z = -\ell$ to $z = 0$, they are *amplified* inside the slab in such a way that their amplitude at the second boundary is a factor $\exp[\kappa_0(2d-\ell)]$ larger than at the source [5]. Similar to the single interface, however, Pendry's solution for evanescent modes is not free of polariton problems. For a slab in vacuum, the polariton dispersion, given by [28]

$$(\kappa - \kappa_0 \varepsilon)/(\kappa + \kappa_0 \varepsilon) = \pm \exp(\kappa d) \quad , \qquad (5)$$

has also the solution $\varepsilon = -1$ for $q \to \infty$. As discussed below, resonant excitation of such modes leads to a divergence of the field for certain intervals of *z*.

To avoid the singularities associated with high-$q$ polaritons, we take $\varepsilon = -1 + \sigma$ (but keep $\mu = -1$) and solve the evanescent-mode problem for an LH-slab in the limit $|\sigma| \ll 1$. The refractive index is $n = -(1-\sigma)^{1/2}$. Note that LH-materials must necessarily exhibit dispersion, i. e., $\sigma$ generally depends on frequency. For calculating the Green's function, the relevant two-dimensional source of $p$-waves is a uniformly distributed line of dipoles which, for simplicity, we place at $z = -d/2$ (the images are at $z = d/2, 3d/2$). The current density is $j_x = p\delta(x)\delta(z+\ell)e^{-i\omega t}$, $j_y = j_z = 0$ and $H(q) = -\text{sign}(z+d/2)p/c$ [11]. Adding (2) and (4), and integrating, we obtain the following expression for the source field

$$H_y^S = \frac{\pi p \omega}{c^2} \times \frac{|z+d/2|}{\sqrt{(z+d/2)^2 + x^2}} H_1^{(1)}\left(\omega\sqrt{(z+d/2)^2 + x^2}/c\right) \qquad (6)$$

containing both propagating and evanescent terms; $H_1^{(1)}$ is a Hankel function. For $z > d$, we write $h = B^- \exp(-\kappa_0 z)$ and use the boundary conditions at $z = 0$ and $z = d$ to obtain $B^-$. Explicitly, for $z > d$ the contribution of evanescent modes to the field is

$$H_y^{NF} = -\frac{p}{c} \int_{|q|>\omega/c} F(q) e^{iqx - \kappa_0 z} dq \qquad (7)$$

where [25]

$$F(q) = \frac{4\kappa\kappa_0 \varepsilon \, e^{\kappa_0 d/2}}{(\kappa_0\varepsilon + \kappa)^2 e^{\kappa d} - (\kappa_0\varepsilon - \kappa)^2 e^{-\kappa d}} \, . \qquad (8)$$

As shown by Pendry [5] using a different method, $F(q) = \exp(3\kappa_0 d/2)$ for $\sigma \equiv 0$. Hence, an ideal slab provides a perfect image of the $q > c/\omega$ components of the source at $z = 3d/2$. By adding the near- and far-field contributions, it can be shown more generally that the total refracted field for $z > 3d/2$ is exactly given by $H_y^S(z - 2d)$. However, notice that $H_y^{NF}$ diverges in the interval $d < z < 3d/2$ if $\sigma \equiv 0$. The limit $\sigma \to 0$ is considered in the following.

Since the singularity is at $q = \infty$, we calculate the field by dividing the integral (7) into two regions: (*i*) $0 < q < Q$ and (*ii*) $q > Q$. Here $Q$ is an auxiliary variable satisfying $\omega/c \ll Q \ll d^{-1} \ln |\sigma|^{-1}$ (the final expression below does not depend on $Q$). In the first region, we set $\sigma = 0$ whereas, in the second region, we deal with the singularity using the approximation $F(q)e^{-\kappa_0 z} \approx e^{-q(z+d/2)}/(e^{-2qd} - \sigma^2/4)$. Details will be discussed elsewhere [30]. Keeping terms of order higher than $\sigma^2$ and replacing $u = z - 3d/2$, we obtain

$$\frac{c}{p} H_y^{NF} \approx \frac{\pi}{2d}\left[\cot[\frac{\pi}{2d}(u-ix)]\left(\frac{\sigma^2}{4}\right)^{(u-ix)/2d} + \cot[\frac{\pi}{2d}(u+ix)]\left(\frac{\sigma^2}{4}\right)^{(u+ix)/2d}\right] \quad (9)$$

$$+ \begin{cases} -2e^{-\frac{\omega u}{c}} \frac{(u\cos\omega x/c - x\sin\omega x/c)}{u^2 + x^2} & u < 0 \\ \pi N_1\left[\omega\sqrt{(u^2+x^2)}/c\right]\frac{\omega u/c}{\sqrt{(u^2+x^2)}} + \int_{-\omega/c}^{+\omega/c}\cos qx \cos[(\omega^2/c^2 - q^2)^{1/2} u] dq & u > 0 \end{cases}$$

where $N_1$ is a Neumann function. A typical field profile is shown in Fig. 1(a) [31]. Consistent with the previous discussion, the real part of the exponent of $\sigma$ is such that, for $\sigma \to 0$, the near field diverges if $u < 0$ while the term that depends on $\sigma$ vanishes if $u > 0$. Accordingly, the length scale of the interference pattern shown in Fig. 1(a) evolves from $d$ for $u < 0$ to $\lambda$ for $u > 0$. Fig. 1(b) is a high resolution image of the region delineated by the rectangle in Fig. 1(a), with the focal point ($x = u = 0$) at its center. The calculated magnetic field, its derivatives and, hence, **E** are all continuous (as they should) at the focal point. We emphasize that (9) is valid for $z > d$ (and, since only terms of order higher than $\sigma^2$ are included [30], for $z < 7d/2$). Using the same procedure, the induced magnetic field can be gained for arbitrary $z$. Inside the slab, i. e., for $0 < z < d$, we get approximately $-\mathrm{sgn}(\sigma)H_y^{\mathrm{NF}}(z+d)+H_y^{\mathrm{NF}}(2d-z)$ whereas, for $z < 0$, we have $-\mathrm{sgn}(\sigma)H_y^{\mathrm{NF}}(-z+d)$ [30]. Here, $H_y^{\mathrm{NF}}(z)$ is the field for $z > d$ as defined in (9). It follows that, depending on the sign of $\sigma$, the induced field is either odd ($\sigma > 0$) or even ($\sigma < 0$) with respect to the center of the slab (thus, there is a second image at the position of the source and, for $\sigma < 0$, a third one at the middle of the lens). The two solutions are shown in Fig. 1 (c) for $x = 0$. This result is not unexpected since the polariton dispersion, Eq. (5), exhibits two branches for which the associated fields have a well-defined parity (the symmetric and antisymmetric functions correspond, respectively, to the low- and high- frequency solutions). These findings are consistent with the time-domain studies of Gómez-Santos [24]. For a time-varying perturbation with a spectrum that is symmetric and centered at the frequency $\Omega$ for which $\sigma \equiv 0$, only the interface at $z = d$ becomes excited due to cancellation between the odd and even solutions; see Fig. 1(c). Within this context, we further note that at the interfaces, where the field is largest, $H_y^{\mathrm{NF}} \propto |\sigma|^{-1/2}$. This is an important result. Since $\sigma \propto (\omega-\Omega)$, this shows that the field induced by a *non-monochromatic* source, of arbitrary frequency spectrum, is an integrable function of $\omega$ for all $x$ and $z$.

The behavior of the total field at the image plane is of particular interest. Adding the radiative component, for which the corresponding expression is $\frac{c}{p}H_y^{\mathrm{R}} = -\int_{-\omega/c}^{+\omega/c}\cos qx \exp[i(\omega^2/c^2 - q^2)^{1/2}u]dq$, we have at $u = 0$

Imaging with an Almost Perfect Lens                                                          R. Merlin

$$\frac{c}{p}(H_y^{NF} + H_y^{R}) \approx \frac{\pi}{d}\coth(\frac{x\pi}{2d})\sin\left[\frac{x}{2d}\ln(\sigma^2/4)\right]. \quad (10)$$

Accordingly, the resolution enhancement is

$$R = -\frac{\lambda}{2\pi d}\ln\left|\frac{\sigma}{2}\right| \quad (11)$$

This expression is identical to that obtained by Smith et al. [25] using a back-of-the-envelope argument, and is also consistent with the analysis of Gómez-Santos [24]. Further, Eq. (10) supports Pendry's claim of perfect imaging in that $\frac{c}{p}(H_y^{NF} + H_y^{R}) \approx -4\pi\delta(x)$ for $\sigma \to 0$. However, as already noted in [25], the resolution is severely limited by the logarithmic dependence of (11) and, moreover, by the fact that the field exhibits a saddle point at $x = z = 0$ so that the depth of focus is poorly defined; see Fig. 1(b). Finally, we observe that the interference pattern with sub-$\lambda$ features described by (10) persists for $u < 0$. This property may be useful for certain applications.

This work was supported by the AFOSR under contract F49620-00-1-0328 through the MURI program and by the NSF Focus Physics Frontier Center. The author would like to thank J. Wahlstrand for help with the calculations and G. Shvets for useful discussions.

FIGURE CAPTIONS

FIG. 1 (color). (a) Calculated contour plot of the magnitude of the magnetic near field, $\mathbf{H}^{NF}$ (logarithmic scale), as a function of $u$ and $x$, in units of $\lambda/2\pi$. The focal point is at $u = x = 0$. Parameters are $\sigma = 10^{-3}$ and $d = \lambda/5\pi$. The green arrow indicates the slab-vacuum interface. (b) Higher resolution image of the near focal point region defined by the green rectangle. Notice that the scale for $|\mathbf{H}^{NF}|$ is linear. (c) Dependence of the field on the direction perpendicular to the slab at $x = 0$. The top (antisymmetric) and bottom (symmetric) solutions correspond, respectively, to positive and negative $\sigma$. The slab, of thickness $d$, is represented by the red rectangle.

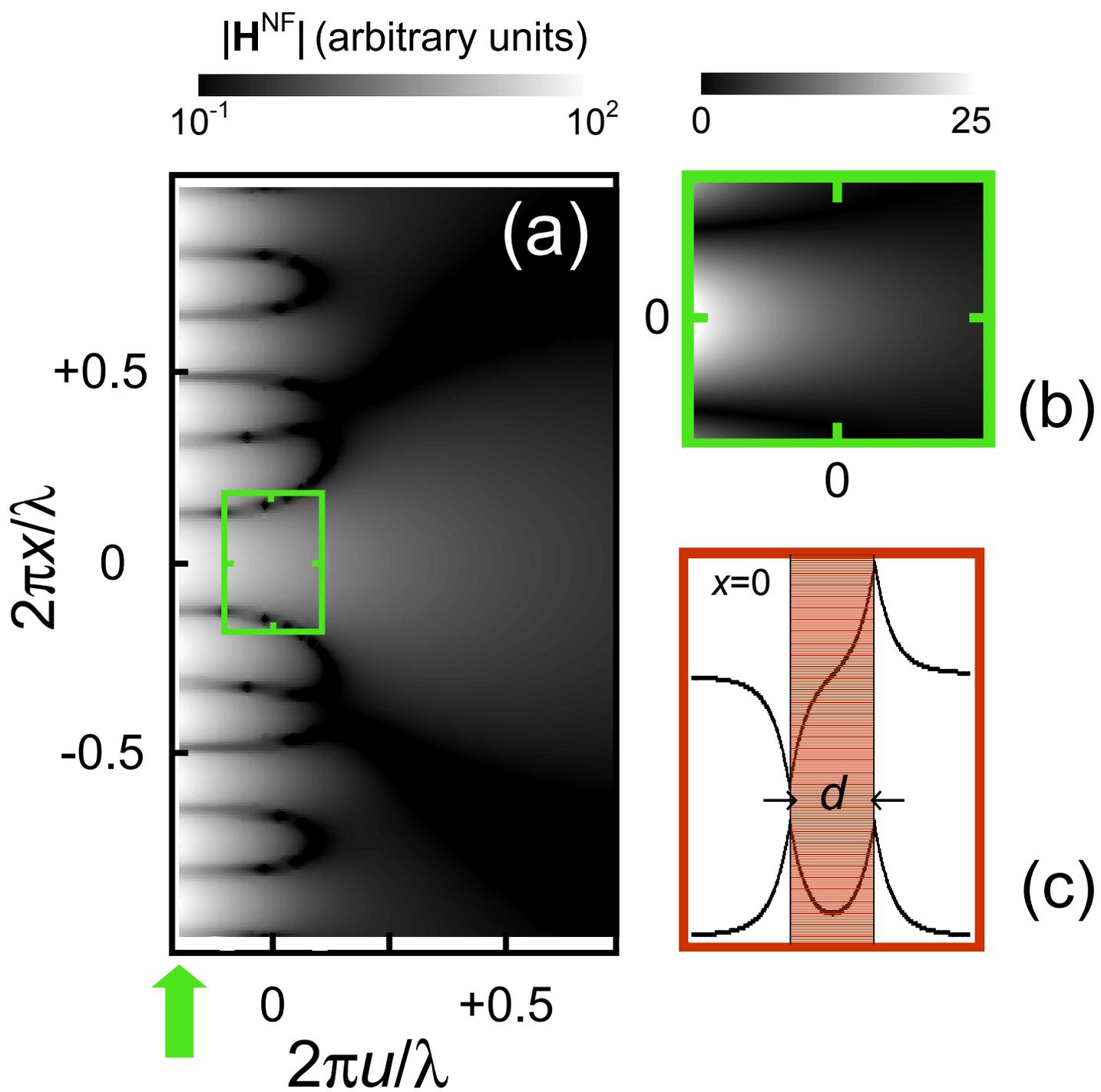